\documentclass[11pt]{article}

\usepackage[a4paper, top=2.5cm, bottom=2.5cm, left=2.5cm, right=2.5cm]{geometry}

\usepackage[sc,osf]{mathpazo}
\usepackage{amsmath,amssymb}
\usepackage{graphicx}
\usepackage{booktabs}
\usepackage{xcolor}
\usepackage{tikz}
\usetikzlibrary{arrows.meta, positioning, calc}
\usepackage{hyperref}
\usepackage{natbib}
\usepackage{microtype}
\usepackage{enumitem}
\usepackage[font=small, labelfont=bf, labelsep=period]{caption}
\usepackage{fontawesome5}
\usepackage[compact]{titlesec}
\titleformat{\section}{\normalfont\large\bfseries}{\thesection}{1em}{}
\titleformat{\subsection}{\normalfont\normalsize\bfseries}{\thesubsection}{1em}{}
\titleformat{\paragraph}[runin]{\normalfont\normalsize\bfseries}{}{0em}{}[.]

\definecolor{accentblue}{HTML}{005AFF}
\definecolor{accentgreen}{HTML}{38A169}
\definecolor{warnorange}{HTML}{DD6B20}
\definecolor{slategrey}{HTML}{4A5568}

\hypersetup{colorlinks=true, linkcolor=accentblue, citecolor=accentblue,
            urlcolor=accentblue}

\setlist{nosep, leftmargin=1.2em}

\begin{document}

\title{\textsc{Triage}: Routing Software Engineering Tasks to Cost-Effective LLM Tiers via Code Quality Signals}

\author{Lech Madeyski\\
  Wroclaw University of Science and Technology, Poland\\
  \texttt{lech.madeyski@pwr.edu.pl}}
\date{}

\maketitle

\begin{abstract}
\textbf{Context:}
AI coding agents route every task to a single frontier large language model (LLM), paying premium inference cost even when many tasks are routine.

\textbf{Objectives:}
We propose \textsc{Triage}, a framework that uses code health metrics---indicators of software maintainability---as a routing signal to assign each task to the cheapest model tier whose output passes the same verification gate as the expensive model.

\textbf{Methods:}
\textsc{Triage} defines three capability tiers (light, standard, heavy---mirroring, e.g., Haiku, Sonnet, Opus) and routes tasks based on pre-computed code health sub-factors and task metadata.
We design an evaluation comparing three routing policies on SWE-bench Lite (300 tasks across three model tiers): heuristic thresholds, a trained ML classifier, and a perfect-hindsight oracle.

\textbf{Results:}
We analytically derived two falsifiable conditions under which the tier-dependent asymmetry (medium LLMs benefit from clean code while frontier models do not) yields cost-effective routing: the light-tier pass rate on healthy code must exceed the inter-tier cost ratio, and code health must discriminate the required model tier with at least a small effect size ($\hat{p} \geq 0.56$).

\textbf{Conclusion:}
\textsc{Triage} transforms a diagnostic code quality metric into an actionable model-selection signal.
We present a rigorous evaluation protocol to test the cost--quality trade-off and identify which code health sub-factors drive routing decisions.
\end{abstract}


\section{Introduction}
\label{sec:intro}

Autonomous coding agents resolve task descriptions by orchestrating planning, code generation, and testing steps.
A critical but under-studied design decision is \emph{which model} should handle each \emph{task} (e.g., issue or feature request).
Using a single frontier model for every task means paying premium cost even when many tasks target clean, well-structured code where a cheaper model would suffice.

Recent work explores multi-model routing at token and query granularities (Section~\ref{sec:related}), but existing approaches suffer compounding errors or rely on model self-confidence and ignore software engineering (SE)-specific metadata.

A natural but unexplored granularity for SE is the coarser \emph{task-level} routing that selects a model tier for an entire task before generation begins.
This granularity is well-suited to SE: task metadata is rich (e.g., file properties, dependencies), extractable from build infrastructure, and domain-specific; and any workflow with a \emph{verification gate} (test suite, linter, type checker) can verify whether the cheaper model succeeded.

The missing piece is a signal that tells the router \emph{when cheap is safe}.
We hypothesize that \citet{Borg26} provide this signal: for single-file refactoring, clean code (\texttt{CodeHealth}~\citep{Tornhill22}~$\geq$~9) yields 15--30\% relative break-rate reductions for medium-sized LLMs, while agentic Claude Code shows no significant difference---the tier-dependent asymmetry a router can exploit.

We propose \textsc{Triage}---a framework that repurposes code health metrics as operational model-selection signals, borrowing the clinical triage principle of matching severity to the appropriate level of care (Section~\ref{sec:framework}).

\textbf{Contributions.}
(1)~We propose task-level model routing for SE: a tier per task/issue granularity that exploits pre-computed, domain-specific metadata unavailable to existing token- or query-level routers.
(2)~We reinterpret the tier-dependent asymmetry of \citet{Borg26} (clean code tolerates cheaper models, messy code does not) as a routing signal, shifting code health from a post-hoc diagnostic to a pre-generation decision criterion.
(3)~\textsc{Triage} framework is abstract on both axes (capability tiers, not named models; any quality indicator, not a proprietary metric), enabling adoption across toolchains.
(4)~We design a reusable, falsifiable evaluation protocol on SWE-bench Lite.

\section{Related Work}
\label{sec:related}

\paragraph{Multi-model routing}
\citet{Chen25} survey LLM--SLM collaboration modes (pipeline, routing, auxiliary, distillation, fusion) and identify task allocation as an open challenge.
RouteLLM~\citep{Ong24} trains a router using preference data to direct queries to a strong or weak model.
Hybrid-LLM~\citep{Ding24} routes based on query difficulty.
Both operate at the query/step level, relying on model self-confidence or preference signals rather than domain-specific metadata.
FusionRoute~\citep{Xiong26} advances token-level collaborative decoding, but compounding errors across sequential token decisions limit its worst-case guarantees.
\textsc{Triage} introduces \emph{task-level} routing, where structured SE metadata provides routing signals unavailable at finer granularities.

\paragraph{AI coding agents}
SWE-bench~\citep{Jimenez24} established the benchmark for agents resolving real GitHub issues.
Current agents default to a single model for every task.

\paragraph{Code health and model performance}
\texttt{CodeHealth}~\citep{Tornhill22} aggregates 25+ sub-factors (e.g., cyclomatic complexity, coupling, file size, duplication, naming) into a 1--10 composite score.
\citet{Borg26} demonstrate that high \texttt{CodeHealth} reduces refactoring break rates for medium-sized LLMs while agentic Claude Code shows no significant difference.

\section{The \textsc{Triage} Framework}
\label{sec:framework}

\subsection{The Core Asymmetry}

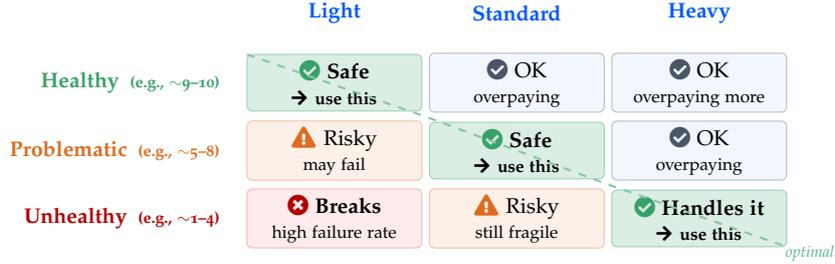
\begin{figure}[t]
  \centering
  \begin{tikzpicture}[
    font=\fontsize{8}{10}\selectfont,
    hdr/.style={font=\fontsize{8}{10}\selectfont\bfseries, inner sep=1.5pt, align=center},
    cell/.style={rounded corners=2pt, minimum width=2.3cm, minimum height=0.75cm,
                 align=center, inner sep=2pt},
    diag/.style={cell, fill=accentgreen!18, draw=accentgreen!50},
    over/.style={cell, fill=accentblue!5, draw=slategrey!40},
    risk/.style={cell, fill=warnorange!10, draw=warnorange!40},
    fail/.style={cell, fill=red!8, draw=red!30},
  ]
    \node[hdr, color=accentblue] at (2.7, 2.05) {Light};
    \node[hdr, color=accentblue] at (5.1, 2.05) {Standard};
    \node[hdr, color=accentblue] at (7.5, 2.05) {Heavy};
    \node[hdr, color=accentgreen, anchor=east] at (1.25, 1.125)
      {Healthy\enskip{\fontsize{6}{7.5}\selectfont(e.g., $\sim$9--10)}};
    \node[hdr, color=warnorange, anchor=east] at (1.25, 0.225)
      {Problematic\enskip{\fontsize{6}{7.5}\selectfont(e.g., $\sim$5--8)}};
    \node[hdr, color=red!70!black, anchor=east] at (1.25,-0.675)
      {Unhealthy\enskip{\fontsize{6}{7.5}\selectfont(e.g., $\sim$1--4)}};
    \node[diag] at (2.7, 1.125)
      {\textcolor{accentgreen}{\faCheckCircle} \textbf{Safe}\\[-1pt]
       {\fontsize{6.5}{8}\selectfont\bfseries\faArrowRight\, use this}};
    \node[over] at (5.1, 1.125)
      {\textcolor{slategrey}{\faCheckCircle} OK\\[-1pt]
       {\fontsize{6.5}{8}\selectfont overpaying}};
    \node[over] at (7.5, 1.125)
      {\textcolor{slategrey}{\faCheckCircle} OK\\[-1pt]
       {\fontsize{6.5}{8}\selectfont overpaying more}};
    \node[risk] at (2.7, 0.225)
      {\textcolor{warnorange}{\faExclamationTriangle} Risky\\[-1pt]
       {\fontsize{6.5}{8}\selectfont may fail}};
    \node[diag] at (5.1, 0.225)
      {\textcolor{accentgreen}{\faCheckCircle} \textbf{Safe}\\[-1pt]
       {\fontsize{6.5}{8}\selectfont\bfseries\faArrowRight\, use this}};
    \node[over] at (7.5, 0.225)
      {\textcolor{slategrey}{\faCheckCircle} OK\\[-1pt]
       {\fontsize{6.5}{8}\selectfont overpaying}};
    \node[fail] at (2.7,-0.675)
      {\textcolor{red!70!black}{\faTimesCircle} \textbf{Breaks}\\[-1pt]
       {\fontsize{6.5}{8}\selectfont high failure rate}};
    \node[risk] at (5.1,-0.675)
      {\textcolor{warnorange}{\faExclamationTriangle} Risky\\[-1pt]
       {\fontsize{6.5}{8}\selectfont still fragile}};
    \node[diag] at (7.5,-0.675)
      {\textcolor{accentgreen}{\faCheckCircle} \textbf{Handles it}\\[-1pt]
       {\fontsize{6.5}{8}\selectfont\bfseries\faArrowRight\, use this}};
    \draw[accentgreen!60, thick, dashed, rounded corners=4pt]
      (1.55, 1.5) -- (3.9, 0.675) -- (6.3, -0.225) -- (8.65, -1.05);
    \node[font=\fontsize{6}{7.5}\selectfont\itshape, color=accentgreen!80,
          anchor=north west] at (8.50, -0.9) {optimal};
  \end{tikzpicture}
  \caption{\textsc{Triage} matrix: code health$\times$model capability tiers.}
  \label{fig:matrix}
\end{figure}

\textsc{Triage} rests on an empirically observed asymmetry: clean, well-structured code can be modified correctly by cheaper models, while messy, complex code requires the reasoning capacity of a frontier model.
While \citet{Borg26} demonstrated this for single-file refactoring, we hypothesize (H1) that the principle extends to multi-step workflows.
The mechanism is structural, not task-specific: clean code reduces the reasoning steps needed to trace dependencies and side effects (a bottleneck for smaller models) while frontier models have excess capacity that absorbs this complexity.
For multi-file tasks, we assume routing is governed by the worst-health file touched, since one complex file can cascade failures through the workflow.
Figure~\ref{fig:matrix} illustrates this as a triage matrix.

\textsc{Triage} places each task near the diagonal, erring above (overpay) rather than below (risk failure) when uncertain.

\subsection{Architecture}

The \textsc{Triage} pipeline operates in three stages:

\paragraph{1. Feature pre-computation}
Per-file code health metrics are pre-computed and incrementally updated (for changed files only) after each task passes the verification gate.
\texttt{CodeHealth} sub-factors (25+) and test coverage are stored in a feature table that the router queries with negligible per-task latency.

\paragraph{2. Routing decision}
Given a task description and the files it references, \textsc{Triage} retrieves pre-computed features and makes a single routing decision before generation begins.
The framework supports three policy families:
\begin{itemize}
  \item \textbf{Heuristic thresholds:} hand-designed rules (e.g., ``\texttt{CodeHealth}~$\geq$~9 $\rightarrow$ light tier''), requiring no training data.
  \item \textbf{Trained ML classifier:} a model learning the feature-to-tier mapping from oracle labels.
  \item \textbf{Perfect-hindsight oracle:} run every task on all tiers, pick the cheapest that passed---an upper bound on routing quality.
\end{itemize}

\paragraph{3. Verification gate}
After the selected model produces its output, the verification gate (test suite, linter, type checker) produces a binary pass/fail signal.
This signal serves two purposes: it validates the current routing decision and, over time, feeds data back to the trained classifier.
A wrong decision is caught and the task is re-executed on the heavy tier.
For per-task costs $c_L < c_S < c_H$ (light, standard, heavy),
let $r_L, r_S$ be the fractions routed to lighter tiers
and $f_L, f_S$ the misrouting rates
(probability that a routed task fails and falls back to heavy).
The expected cost per task is
\begin{equation}\label{eq:cost}
  r_L\!\left(c_L + f_L\, c_H\right)
  + r_S\!\left(c_S + f_S\, c_H\right)
  + \left(1 - r_L - r_S\right) c_H\,,
\end{equation}
saving $r_L(c_H{-}c_L) + r_S(c_H{-}c_S)
       - (r_L f_L {+} r_S f_S)\,c_H$
per task over always-heavy:
the perfect-routing margin minus a fallback penalty.
In the two-tier case this simplifies to
pass rate $1{-}f_L > c_L/c_H$,
recovering the cost gate of Section~\ref{sec:evaluation}.
Feature pre-computation is amortized across tasks; incremental updates are negligible.

\paragraph{Dual abstraction}
\textsc{Triage} is abstract on both axes: it routes to capability \emph{tiers} (not named models), so swapping the underlying model requires only tier--model recalibration; and it accepts any per-file quality indicator, not a single proprietary metric.

\section{Proposed Evaluation}
\label{sec:evaluation}

\paragraph{Hypotheses}
\textbf{H1:} The tier-dependent asymmetry extends to multi-step issue resolution on SWE-bench Lite.\\
\textbf{H2:} \texttt{CodeHealth} discriminates the required model tier with at least a small effect size ($\hat{p} \geq 0.56$).

\paragraph{Dataset}
SWE-bench Lite~\citep{Jimenez24} provides 300 real GitHub issue-resolution tasks of varied difficulty.
Each task is run on all three tiers, three times per tier for non-determinism handling (majority-vote pass rule), yielding 2,700 agent runs (evaluation only; deployment runs once per Eq.~\ref{eq:cost}).
To test generalizability, both open-weight and cloud-hosted model families are evaluated.

\paragraph{Matched-pair design}
For analysis, tasks are paired on difficulty proxies (e.g., patch size) to isolate the code health signal from task difficulty.

\paragraph{Policies compared}
The three policy families (Section~\ref{sec:framework}) are evaluated against three baselines: always-light, always-heavy, and random.

\paragraph{Feature importance (exploratory)}
Contingent on H1--H2, we investigate \textbf{RQ1:} \emph{Does the composite \texttt{CodeHealth} score add routing value beyond its individual sub-factors?}
RQ1 reuses the same oracle labels (no additional agent runs are needed; only lightweight classifiers are retrained): SHAP analysis~\citep{Lundberg17} on a training split ranks sub-factors; on a held-out split, classifiers using the top-1, top-3, and top-5 sub-factors are compared against the composite on MCC and cost savings (Eq.~\ref{eq:cost}).

\paragraph{Metrics}
Task success rate, cost per successful task, triage accuracy (vs.\ oracle), over-triage and under-triage rates.
Results are stratified by test coverage of changed code.
Statistical methods: Brunner-Munzel test and the probability of superiority ($\hat{p}$) effect size.

\paragraph{Pilot go/no-go criteria}
A 50-task pilot tests \textsc{Triage}'s core asymmetry on the extreme tiers (light vs.\ heavy), where the capability gap maximizes the detectable effect size, before committing to the full three-tier, 300-task evaluation.
Following \citeauthor{Kitchenham24}'s~\citeyearpar{Kitchenham24} recommendation to prefer effect sizes over significance tests for small samples, the go/no-go decision requires \emph{both}:
\begin{itemize}
  \item \emph{Cost gate:} a try-light-first strategy with heavy-tier fallback must be cheaper than always-heavy, i.e., the light tier's pass rate on routed tasks must exceed the cost ratio $c_L/c_H$ (e.g., 20\% for Haiku$\to$Opus at current API pricing).
  \item \emph{Signal gate:} the probability of superiority $\hat{p}$ for high- vs.\ low-\texttt{CodeHealth} tasks must reach at least a small effect ($\hat{p} \geq 0.56$~\citep{Kitchenham24}), confirming that code health discriminates task difficulty.
\end{itemize}
Both gates are needed: the cost gate alone cannot confirm that code health drives the savings; the signal gate alone cannot confirm the savings outweigh failures.
If either fails, the negative result is reported as-is.

\section{Discussion}
\label{sec:discussion}

\textsc{Triage}'s practical value depends on the strength of the code health signal.
If the core asymmetry holds, the frontier-to-light cost ratio defines the savings ceiling.
The construct validity check is methodologically critical: if individual sub-factors predict the required tier, the composite score is unnecessary and \textsc{Triage} works with freely available metrics, lowering adoption barriers.
In Spec-Driven Development workflows, routing savings apply to every task.
In evaluation, target files are known from ground-truth patches; in deployment, they must be inferred from the issue description.
Issue-driven workflows also benefit, though the signal weakens with incomplete test coverage.
Three validity threats deserve discussion: (i) code health and task difficulty may confound (files with low health scores often accompany harder tasks) making the matched-pair design essential to isolate the routing signal; (ii) \texttt{CodeHealth} is a proprietary metric, so the construct validity check against individual sub-factors is critical for replicability; (iii) in deployment, issue descriptions may not name target files, requiring heuristics whose accuracy bounds \textsc{Triage}'s practical effectiveness.

A further direction is \emph{(sub-task/step)-level} routing: assigning tiers to agent steps within a task/issue.
Task-level routing is necessary to establish whether code health discriminates difficulty. Sub-task routing adds cross-step coherence challenges that we leave to future work.

This paper presents a new idea, not yet fully proven.
We contribute (i)~task-level routing guided by code health---reinterpreting a diagnostic metric as a decision signal, (ii)~a reusable, falsifiable evaluation protocol, and (iii)~a framework abstract on model and metric axes.
A negative result (code health failing to discriminate model-tier requirements) would be equally informative, narrowing the search for useful routing signals in SE.

\end{document}